\documentclass[envcountsame]{llncs}
\usepackage{color} 
\usepackage{graphicx}

\newtheorem{thm}{Theorem}

  \renewcommand{\theenumi}{\roman{enumi}}

\def \qedbox{\hfill\vbox{\hrule\hbox{\vrule height1.3ex\hskip0.8ex\vrule}\hrule}}
  \newfont{\MSB}{msbm10 scaled\magstep1}

\definecolor{grey}{gray}{0.85}

\begin{document}

\title{Adversarial Scheduling in Evolutionary Game Dynamics}

\author{Gabriel Istrate\inst{1}\thanks{corresponding author. Email: gabrielistrate@acm.org}\and Madhav V.~Marathe \inst{2} \and S.S.~Ravi\inst{3}}
\institute{e-Austria Institute, V.P\^{a}rvan 4, cam. 045B, Timi\c{s}oara RO-300223, Romania \and Network Dynamics and Simulation Science Laboratory,
Virginia Tech, 1880 Pratt Drive Building XV, Blacksburg, VA 24061. Email: \email{mmarathe@vbi.vt.edu} \and Computer Science Dept., S.U.N.Y. at Albany,
  Albany, NY 12222, U.S.A. Email:\email{ravi@cs.albany.edu}}

\maketitle




\begin{abstract}
Consider a system in which players at nodes of an underlying graph
$G$ repeatedly play Prisoner's Dilemma against their neighbors.  The
players adapt their strategies based on the past behavior of their
opponents by applying the so-called win-stay lose-shift strategy. This
dynamics has been studied in \cite{kittock:colearn,ipd:colearning,mossel-ipd}.

With random scheduling, 
starting from any initial configuration with high probability the
system reaches the unique fixed point in which all players cooperate.
This paper investigates the validity of this result under various
classes of {\em adversarial schedulers}. Our results can be summarized
as follows:
\begin{itemize}
\item
An adversarial scheduler that can select {\em both} participants to the game
can preclude the system from reaching the unique fixed point on most
graph topologies.
\item A nonadaptive scheduler that is only allowed to
 choose {\em one} of the participants is no more powerful than a
 random scheduler. With this restriction even an adaptive scheduler is not significantly
 more powerful than the random scheduler, provided it is ``reasonably fair''.
\end{itemize}

The results exemplify the {\em adversarial scheduling}
approach we propose as a foundational basis for the {\em generative
approach to social science} \cite{epstein-generative-book}.

\end{abstract}


\baselineskip=1.05\normalbaselineskip

\section{Introduction}

Evolutionary game theory \cite{weibull-egt} and agent-based
simulation in the social sciences
\cite{ballot-weisbuch-acs,ace-handbook,gilbert-troizch} 
share the object of study and a significant set of
concerns, and are largely distinguished only by the method of choice
(mathematical reasoning versus computational experiments). In
particular, the two areas deal with fairly similar models (see, e.g. \cite{wilhite-handbook,BEM05}).  A particular class of such models assumes a large population of agents located at the vertices of a graph. Agents interact by playing a fixed game, and update their behavior based on the outcome of this interaction (according to a pre-specified rule). 

How does one interpret properties of such systems, be they obtained
through mathematical analysis or computational simulations ? A possible answer is that results characterizing dynamical properties of such models provide insights (and possible explanations) for features observed in ``real-world'' social dynamics. For instance, the primary intuition behind the concept of {\em stochastically stable strategies} in evolutionary game theory is that a small amount of ``noise'' (or, equivalently, small deviations from rationality) in game dynamics can solve the
equilibrium selection problem, by focusing the system
on one particular equilibrium. Similarly, in agent-based social
 theory, Epstein \cite{epstein-generative-book,epstein-generative} (see also
\cite{axtell-epstein}) has advocated a generative approach to social
science. The goal is to {\em explain} a given social phenomenon by {\em
generating} it using multiagent simulations\footnote{cf.
\cite{epstein-generative-book}, Chapter 1: ``if you didn't grow it
[the social phenomenon, n.n.], you didn't explain its emergence''.}.
Related concerns have been recently voiced throughout analytical
social science, with a particular emphasis on {\em mechanism-based
explanations} \cite{hedstrom-dissecting,hedstrom-social}.

Given that game theory and agent-based simulation are emerging as tools for guiding political decision-making (see e.g. \cite{BEM05,epstein-smallpox,EG+04,transims}), it is important to
make sure that the conclusions that we derive from these techniques are {\em robust} to small variations in
model specification\footnote{The importance of stability has been
previously recognized in the literature, for instance, in \cite{axtell-epstein} (pp. 35), where emergent phenomena are defined as ``{\em stable} macroscopic patterns arising from local of interaction of agents''.}.

In this paper we consider the effect of one particular factor that can affect
the robustness of results of agent-based simulations and evolutionary
game theory: {\em agent scheduling}, i.e. the order in which agents
get to update their strategies.  Many models in the literature assume one of two popular alternatives:
\begin{itemize} 
 \item {\bf synchronous update:} {\em every}
player can update at every moment (in either discrete or continuous
time); this is the model implicitly used by ``large population"
models in evolutionary game theory.
\item {\bf uniform matching:} agents are vertices of a 
(hyper)graph.  At each step we choose a (hyper)edge uniformly 
at random and all players corresponding to this hyperedge are 
updated using the local update function.
\end{itemize}

Instead of postulating one of these two update mechanisms, we 
advocate the study of social dynamics under an approach we call
{\em adversarial scheduling}.  We exemplify
 adversarial scheduling by studying, in such a setting, the 
Iterated Prisoners' Dilemma game with win-stay lose-shift 
strategy. This dynamics (originally motivated by the 
colearning model in \cite{shoham:tenn:colearn:aij}) have 
received substantial attention in the game-theoretic 
literature\cite{kittock:colearn,ipd:colearning,mossel-ipd}.

We are, of course, far from being the first researchers to 
recognize the crucial role of scheduling/activation order on 
the properties of social dynamics (to only give two examples, 
see \cite{huberman:glance,axtell00effects}). However, what we 
advocate is a  (somewhat) more systematic approach, based on 
the following principles:
\begin{enumerate}
\item {\em Start with a ''base case`` result $P$, stated under 
a particular scheduling model.}
\item {\em Identify several structural properties of the scheduling model that impact the validity of $P$.}
Ideally, these properties  should be selected by careful examination of the proof of $P$,
which should reveal their importance.
\item {\em Identify those properties (or combinations of 
properties) that are {\bf necessary/sufficient} for the 
validity of $P$.} Correspondingly, {\em identify properties 
that that are inconsequential to the validity of $P$.}
\item The process outlined so far can continue by recursively 
applying steps (i)-(iii). In the process {\em we may need to  
reformulate the original statement in a way that makes it hold 
under larger classes of schedulers}, thus making it more 
robust. The precise reformulation(s) normally arise from 
inspecting the cases when the proof of $P$  fails in an 
adversarial setting.
\end{enumerate}

The intended benefits of the adversarial scheduling approach 
are multiple (see \cite{adversarial-jasss} for further discussions). 
The aim of this paper is to show that these benefits do not come  at the expense of mathematical tractability: at least for one  nontrivial dynamics, adversarial scheduling (as outlined in the five point approach above) is feasible and can
lead to interesting results.

\section{Preliminaries}

First we define two classes of graphs that we will frequently
consider in this paper. The {\bf line graph} $L_n$, $n \geq 1$
consists of $n$ vertices $v_1, \ldots v_n$ and edges $(v_i, v_{i+1})$,
$1 \leq i \leq n-1$. We will use {\em Line} to denote the set of all
line graphs.  A {\bf star graph} $Star_n$, $n \geq 1$, is the complete
bipartite graph $K_{1,n}$.  We will use {\em Star} to denote the set
of all star graphs.

\subsection{Basic Model: Prisoner's Dilemma with Pavlov dynamics}

Next we describe the basic mathematical model for Prisoner's Dilemma with
Pavlov dynamics; see \cite{kittock:colearn,mossel-ipd,ipd:colearning}
for additional discussion.  We are given an undirected graph $G(V,E)$,
$|V|= n$ and $|E| = m$.  Each vertex $v \in V$ represents an agent.
Each agent has a label from the set $\{0,1\}$. These labels denote the
strategies that the players follow: $0$ can also be equivalently
viewed as {\em cooperation} and $1$ can be viewed as {\em defection}.
Without loss of generality, we assume that $G$ is connected, otherwise
the dynamics will reduce to independent dynamics on the connected
components of $G$. We will also assume that the graph contains at
least two edges.

Time will be discrete. At time $t =0$ all nodes are assigned a
label from $\{0,1\}$. At each subsequent step, certain nodes/agents
change their label (strategy) according to the rules given below. We
will use $x_t(v)$ to denote the label of node $v$.
At each step $t+1$ an edge $e = (u,v)$ is
selected according to some rule and the states of $u$ and $v$ are
updated as follows:
\[x_{(t+1)}(u)  \leftarrow  (x_t(u) + x_t(v))~(\textrm{mod}~2)\]
\[x_{(t+1)}(v)  \leftarrow  (x_t(u) + x_t(v))~(\textrm{mod}~2)\]
We will use ${\bf X}_t$ to denote the vector $(x_t(v_1), \ldots
x_t(v_n))$ representing the states of the nodes $v_1, \ldots
v_n$. This will be sometimes referred to as the {\em global
configuration}.  Sometime we will omit the subscript $t$ for ease of
exposition and its value will be clear from context.  With this
terminology, each step of the dynamics can be viewed as a global
update function $F$. It takes as input an element $e= (v_i, v_j) \in
E$, ${\bf X} = (x(v_1), \ldots x(v_n))$ and returns the next global
${\bf Y} = (y(v_1), \ldots y(v_n))$ given as follows: $\forall v_k$,
s.t. $k \neq i$ and $k \neq j$, $y(v_k) = x(v_k)$; and $y(v_i)
= y(v_j) = (x(v_i) + x(v_j))~(\textrm{mod}~2)$.  In this case ${\bf
Y}$ is said to be reachable from ${\bf X}$ in one step.  A global
configuration ${\bf X}$ is said to be a fixed point if $\forall e \in
E$, $F({\bf X}, e) = {\bf X}$. It is easy to see the dynamical system
studied here has a unique fixed point ${\bf 0} = (0, \ldots,
0)$. Following dynamical systems literature, a configuration ${\bf X}$
is called a {\em Garden of Eden configuration} if the configuration is not
reachable from any other configuration.  For the rest of this paper,
we will use ${\bf X}, {\bf Y}, \ldots$ to denote global
configurations.  An instance of the Prisoner's Dilemma with Pavlov
dynamics (PDPD) can thus be represented as a $(G,f)$, where $G$ is the
underlying interaction graph and $f$ is the local function associated
with each node. In the remainder of this paper, since $f$ is always
fixed, an instance will be specified simply by $G$.

\subsection{The base-case result}

The following property of the PDPD is easily seen to hold under random matching: for all interaction graphs $G$ with no isolated vertices {\em the system converges with probability $1-o(1)$  to
 the ``all zeros'' configuration (henceforth denoted ${\bf 0}$) }  With a slight abuse of convention, we will refer to this event as {\em self-stabilization}. This will the statement we will aim to study in an adversarial setting.

We will also be interested in the {\em convergence time} of the
dynamics. Under random scheduling Dyer et al. \cite{ipd:colearning}
prove that the number of steps needed to self-stabilize is $O(n\log
n)$ on $C_n$ (the simple cycle on $n$ nodes) and exponential in $n$ on
$K_n$ (the complete graph on $n$ nodes).  The convergence time was
further investigated by Mossel and Roch \cite{mossel-ipd}.

\subsection{Types of scheduler}

 A {\em schedule} ${\bf S}$ is specified as an infinite string over $E$, i.e.
${\bf S} \in E^*$. Given a schedule ${\bf S} = (e_1, e_2, \ldots, e_t, \ldots)$, the
graph $G$ and an initial configuration ${\bf I}$, the dynamics of the
system evolve as follows. At time $t = 0$ the system is in state ${\bf
I}$. At time $t$, we pick the $t^{th}$ edge from ${\cal S}$. Call this
edge $e_t$. If the configuration at the beginning of time $t$ is ${\bf
X}$ then the configuration ${\bf Y}$ at the beginning of time $t+1$ is
given by ${\bf Y} \leftarrow F({\bf X}, e_t)$. The iterated global
transition function $F^*$ is defined as
\[ F^*({\bf I}, (e_1, e_2, \ldots, e_t, \ldots)) \equiv  F^*(F({\bf I}, e_1), (e_2, \ldots, e_t, \ldots))\]

We say that a system $G$ {\em self stabilizes} for a given initial
configuration ${\bf I}$ and a schedule ${\bf S} = (e_1, \ldots, e_t,
\ldots)$\footnote{Note that $e_i$ and $e_j$ in the sequence need not
be distinct} if \ $\exists t \geq 1$ such that the system starting in
${\bf I}$ reaches the (unique fixed point) configuration ${\bf 0}$
after $t$ time steps, i.e.  ${\bf 0} \leftarrow F({\bf I}, (e_1,
\ldots, e_t))$.  $G$ is said to self stabilize for a schedule $S$ if
$G$ eventually reaches a fixed point when started at any initial
configuration $I$, i.e. $\forall I, \ F({\bf I}, {\bf S}) \rightarrow
{\bf 0}$.  Conversely, a schedule ${\bf S}$ can preclude self
stabilization of $G$ if $\exists I$ such that $F({\bf I}, S)$ does not
ever reach ${\bf 0}$.  Given a set of schedules, a scheduler is simply
an algorithm (possibly randomized) that chooses a schedule. The
schedulers considered here are all polynomial time algorithms and the
set of feasible schedules are described below.

Schedulers can be adaptive or non-adaptive. An adaptive scheduler
decides on the next edge or node based on the current global
configuration.  A non-adaptive scheduler decides on a schedule in
advance by looking at the graph (and possibly the initial
configuration). This schedule is then fixed for remainder of the
dynamic process.  One particularly restricted class of {\em
non-adaptive} schedules is a fixed permutation of nodes/edges repeated
periodically and independent of the initial state of the system.  For
node updates, this is the model employed in {\em sequential dynamical
systems} \cite{BH+03,BEM05}.

There are several distinctions that can be made, concerning the power
of a scheduler. The first one concerns the number of players that the
scheduler is able to choose.  There are two possibilities.

\begin{itemize}
\item An {\em edge-daemon} (or edge-scheduler) is able to choose {\em
both} players of the interacting pair. In other words, an edge daemon
constructs ${\bf S}$ by selecting edges $e_i \in E$ in some order.
\item A {\em node-daemon} (or node-scheduler) can choose
  only one of the players. We can let this player choose its
  partner. A natural model is to consider the case where partner
  is chosen uniformly at random among the neighbors of the
  first player. In such a case, we say that the node-scheduler
  model has {\em random choice}, or random-choice node-scheduler.
\end{itemize}

\subsection{Summary of results}

Our results can be summarized as follows:

\begin{itemize}
\item Not surprisingly, some amount of fairness is a necessary 
 to extend self-stabilization in an adversarial setting. 
\item The power given to the
scheduler makes a big difference in whether or not the system 
self stabilizes:
\begin{enumerate}
\item if the scheduler can {\em exogenously} choose
{\em both} participants in the game then (Theorems 2,3) it can
preclude convergence on most graphs, even when bounded by fairness
constraints.
\item On the other hand schedulers that allow a limited amount of {\em
endogeneity in agent interactions}\footnote{The importance of this
property has been recognized \cite{vriend-handbook} in the agent-based
simulation literature}, by only choosing {\em one} of the participants
are no more powerful than the random scheduler (Theorem 4) when
nonadaptive, and are not significantly more powerful (Theorem 5) when
adaptive but ``reasonably fair''.
\end{enumerate}
\end{itemize}

We also investigate experimentally (in Section~\ref{times-sim}) the
convergence time of the colearning dynamics for a few nonadaptive
schedulers, in a case when the convergence time for the random
scheduler is known rigorously.

\subsection{Related work}

Our approach is naturally related to the {\em theory of self-stabilization of
distributed systems} \cite{dolev:book}.  Multi-agent systems, like the
ones considered in the evolutionary game dynamics, have many of the
characteristics of a distributed system: a number of entities (the
agents) capable of performing certain {\em computations} (changing
their strategies) based on {\em local information}. Randomized models
of this type (including the model we study in this paper) have been in
fact recently considered in the context of self-stabilization
\cite{fribourg-ipd}. There are, however, a number of
differences. First, in self-stabilization the computational entities
(processors) are capable of executing a wide-range of activities
(subject to certain constraints, for example the requirement that all
processors run the same program, in the context of so-called {\em
uniform self-stabilizing systems}). The goal of such systems is to
achieve a certain goal (legal state) in spite of transient errors and
malicious scheduling. In contrast, in our setup, there is no ``goal'',
the computations are fixed, and restricted to steps of the
evolutionary dynamics. The only source of uncertainty arises from the
scheduling model. A second difference is in the nature of the update
rule. Usually, in self-stabilizing systems there is a difference between
{\em enabled} processors, that intend to take a step, and those that indeed
take it. Such a notion is not so natural in the context of game-theoretic models.

As mentioned earlier, the dynamics, can be easily recast in the
context of Prisoner's Dilemma: let $1$ encode ``defection'' and $0$
encode ``cooperation''. Then the update rule corresponds to the
so-called {\em win-stay, lose-shift} \cite{posch:wsls}, or Pavlov
strategy. This rule specifies that agents defect on the next move
precisely when the strategy they used in the last interaction was
different from the strategy used by the other player.  It was the
object of much attention in the context of Iterated Prisoner's Dilemma
\cite{axelrod:book:1},
\cite{nowak:sigmund:pavlov},\cite{axelrod:book:2}. Related versions of
the dynamics have an even longer history in the Psychology literature,
where they were proposed to model emergence of cooperation in
situations where players do not know precisely the payoffs of the game
in which they are participating, and might even be unaware they are
playing a game: Sidowski \cite{sid:minimal:situation} has proposed the
``minimal social situation" (MSS), a two-person experiment
representing an extremely simple form of interaction between two
agents. MSS was first viewed as a game by Thibaut and Kelley
\cite{thibault:kelley} who called it ``Mutual Fate Control".
An explanation for the empirical observations in \cite{thibault:kelley} was
proposed by Kelley, Thibaut, Radloff and Mundy \cite{ktrm}, that
raised the possibility that players were acting according to the
Pavlov dynamics. MSS was generalized to multiplayer games by Colman et
al. \cite{n:person:MSS,colman:grant}, who obtained mathematical
characterization for the emergence of cooperation.

\section{Fairness in scheduling}

A necessary restriction on schedulers we will be concerned with is
{\em fairness}. In self-stabilization this is usually taken to mean
that each node is updated infinitely often in an infinite schedule. We
will also consider notions of {\em bounded fairness}. A natural
definition is the following:

\begin{definition}
Let $b\geq 1$. A scheduler that can choose one item among a set of $m$
elements is {\em (worst-case) $b$-fair} if for every agent $x$,
no other agent is scheduled more than $b$ times between two
consecutive times that $x$ is scheduled.
\end{definition}

It is easy to see that a 1-fair edge scheduler chooses a fixed permutation of edges uses
this as a periodic schedule. A 1-fair node daemon
selects a fixed permutation of nodes and for each node selects a
random neighbor and repeats the same permutation (with possibly
different partners) periodically.

The fact that we want to investigate properties of the random scheduler suggest investigating fairness of probabilistic schedulers. For such schedulers, the worst-case fairness in Definition 1 is far too restrictive. 

\begin{definition} 
 A probabilistic scheduler is {\em weakly fair} if for any node $x$ and any initial schedule $y$, the probability that $x$ will eventually be scheduled, given that the scheduler selected nodes according to $y$ is positive. 
\end{definition}

The random scheduler is weakly fair. Not every scheduler is 
weakly fair, and a scheduler need not be weakly fair to make 
the system self-stabilize. On the other hand, the base-case 
result does not extend to the adversarial setting when weak 
fairness is not required: 

\begin{thm}
The following are true: 
\begin{enumerate} 
\item There exists an edge scheduler that is not weakly fair 
and that makes the system self-stabilize no matter what its 
starting configuration is. 
\item For any graph $G$ there exists an edge/node scheduler 
that is not weakly fair and that prevents the system from self-stabilizing on some initial configuration. 
\end{enumerate} 
\end{thm} 

Another restatement of Theorem 1 is that weak fairness is necessary to preclude some ''degenerate`` schedulers like the ones we construct for the proof of point (ii). 
\begin{proof} 
\begin{enumerate} 
\item Consider an edge scheduler that works as follows: 
\begin{itemize} 
\item Choose an edge $e$ that has not yet self-stabilized, i.e. at least one of its endpoints is ${\bf 1}$. 
\item Turn nodes of $e$ to ${\bf 0}$ by playing $e$ twice.  
 \item The scheduler never schedules $e$ subsequently. 
\end{itemize}
\item Consider a node (edge) scheduler that repeatedly schedules the same node (edge). It is easy to see that the system does not self-stabilize unless the graph consists of a single edge (a star in the case of a node scheduler when the center node is the scheduled one). 
\end{enumerate}
\end{proof}

\begin{definition}

A probabilistic scheduler is {\em $O(f(n))$-node fair w.h.p.} if
the following condition is satisfied while the system has not reached the fixed point: {\em For any schedule $W$ (call its last scheduled node $x$), every node $y$ and every $\epsilon >0$ there exists $C_{\epsilon}>0$ such that, with probability at least $1-\epsilon$ node $y$ will be scheduled at most $C_{\epsilon}\cdot f(n)$ times before $x$ is scheduled again.}. 

A scheduler is {\em boundedly node fair w.h.p.} if it is 
$O(f(n))$-node fair w.h.p. for some function $f(n)$.
\end{definition}

We emphasize the fact that Definition 2 applies to edge 
schedulers as well, when a node $x$ is considered to occur at 
stage $t$ if some edge containing $x$ is scheduled at that 
step. 

With these definitions we have: 

\begin{thm}
Let $S$ be a (node or edge) weakly fair probabilistic 
scheduler such that the following result holds: for any
initial configuration, the probability that the system
self-stabilizes tends to one. Then the scheduler is boundedly node fair. 
\end{thm}

\begin{proof} 
 We will consider both node and edge schedulers at the same time. 
Let $\epsilon >0$ and let $T=T(\epsilon,n)$ be an integer such that, no matter in which configuration $S$ we start the system, the probability that the system does {\em not } self-stabilize (taken over all the coin tosses of the scheduler) is at most  $\epsilon$. 

Consider any state $S$ of the system after a node $x$ has be scheduled, and assume that $S$ is not the absorbing state ${\bf 0}$. Run the system for $T$ steps. The probability that the system 
does not self-stabilize is at most $\epsilon$. On the other hand, if some node $y$ is not played at all during the $T$ steps then 
the system has no chance to self-stabilize. It follows that the maximum number of times a given node can be scheduled before $x$ is scheduled again is at most $T-1$. 
 
\end{proof}

\section{The Power  of Edge-Schedulers}

From now on we will restrict ourselves to boundedly fair 
schedulers. This section aims to show that edge-schedulers are 
too powerful. Indeed, it is easy to show that there exist 
graphs on which even 1-fair edge-schedulers can prevent 
self-stabilization. The following two results provide a modest improvement, showing that even 2-fair edge-daemons on any graph are too strong : 

\begin{thm}
Let $G$ be an instance of PDPD. Then there exists an initial
configuration $I$ and a 2-fair edge-scheduler {\bf S} that precludes
self-stabilization on $G$ starting in configuration $I$.
\end{thm}

\begin{proof}
Consider a sequence of edges
$e_{0}, \ldots e_{k}$ (with repetitions allowed) such that

\begin{itemize}
\item every edge of $G$ appears in the list.

\item for every $i=0, \ldots, k$, $e_{i}$ and $e_{i+1}$ have {\em exactly
one} vertex in common (where $e_{k+1}= e_{0})$.
\item Every edge appears in the sequence at most twice.
\end{itemize}

We will show that an enumeration $F(G)$ with these properties can be
found for {\em any} connected graph with more than one edge. Such a
sequence specifies in a natural (via its periodic extension) a
2-fair  edge daemon. It is easy to see that the only states that can
lead to the fixed point are the fixed point and states leading to it
in one step: such a state has exactly two (adjacent) ones. But such
a state cannot be reached from any other state according to the
previously described scheduler, since the edge that would have been
``touched" immediately before has unequal labels on its extremities,
which cannot be the case after updating it.

We now have to show how to construct the enumeration $F(G)$. First
we give the enumeration in the case graph $G$ is a tree. In this
case we perform a walk on $G$, listing the edges as follows: suppose
the root $r$ is connected, via vertices $v_{1}, \ldots, v_{k}$ to
subtrees $T_{1}, \ldots, T_{k}$. Then define recursively
\begin{equation}\label{2:tree}
F(G)=(v,t_{1})F(T_{1})(t_{1},v)(v,t_{2})F(t_{2})(t_{2},v)\ldots
(t_{k-1},v)(v,t_{k})F(T_{k})(t_{k},v),
\end{equation}

where, if the list of edges thus constructed contains two
consecutive occurrences of the same edge we eliminate the second
occurrence.

Consider now the general case of a connected graph $G$, and let
$S(G)$ be a spanning tree in $G$. Edges of $G$ belong to two
categories:
\begin{enumerate}
\item Edges of the spanning tree $S(G)$.
\item Edges in $E(G)\setminus E(S(G))$.
\end{enumerate}

Define $F(G)$ as the list of edges obtained from $F(S(G))$ in the
following way: whenever $F(S(G))$ first touches a new vertex $w$ in
$G$ insert the edges in $E(G)\setminus E(S(G))$ adjacent to $w$ (in
some arbitrary order); continue then with $F(S(G))$. It is easy to
see that each edge $e$ is listed at most twice in $F(G)$. To prove
this consider the two cases, $e\in E(S(G))$ and $e\in E(G)\setminus
E(S(G))$. The statement follows in the first case by the recursive
definition~\ref{2:tree}. In the second case it follows by
construction, since a nontree edge is visited only when one of its
endpoints is first touched in $E(S(G))$.

\end{proof}

Even the most restricted edge-schedulers, 1-fair edge-schedulers,
are able to preclude self stabilization on a large class of graphs.
To see that define
\begin{enumerate}
\item ${\cal G}_{1}$ to be the class of graphs $G$ that contain a
cycle of length at least four.
\item ${\cal G}_{2}$ to be the class of graphs $G$ that contain {\em
no} cycles of length at least four and $m$, the number of edges of $G$
is even.
\item ${\cal G}_{3}$ to be the class of trees  with $n= 4k$ vertices.
\end{enumerate}

\begin{thm}
Let $G$ be a connected graph in ${\cal G}_{1}\cup {\cal G}_{2}\cup
{\cal G}_{3}$. Then there exists an initial configuration on $G$ and a
{\em 1-fair} edge-schedule {\bf S} that is able to forever preclude
self-stabilization on $G$.
\end{thm}

In other words, connected graphs for which the system self-stabilizes
for all 1-fair schedulers have an odd number of edges and all their
cycles (if any) have length 3.

\begin{proof}
Define $\Delta_{i,j} = (d_{k,l}^{(i,j)})$ a $n\times n$ matrix over ${\bf Z_{2}}$ by

\begin{equation}
d_{k,l}^{(i,j)}=\left \{\begin{array}{ll}
             1  &\mbox{, if }(k,l)=(i,j) \\
             0  &\mbox{, otherwise. }
             \end{array}
      \right.
\end{equation}

Suppose we represent configurations of the system as vectors in ${\bf Z}_{2}^{n}$.

Taking one step of the dynamics on an arbitrary
configuration ${\bf X}$ with the scheduled edge being
$(i,j)$ leads to configuration $\overline{{\bf X}}= A_{i,j}\cdot {\bf X}$,
where matrix $A_{i,j}$ is given by

\begin{equation}\label{first-eq}
A_{i,j}=I_{n}+\Delta_{i,j}+ \Delta_{j,i}.
\end{equation}

Indeed, the only nondiagonal elements of matrix $A_{i,j}$ that are nonzero are in positions
$(i,j)$ and $(j,i)$. This means that all elements of a configuration
${\bf X}$ in positions other than $i,j$ are
preserved under multiplication with $A_{i,j}$. It is easy to see that labels in positions $i,j$ change according to the specified dynamics.

Consider a graph $G$ with $m$ edges,
$E(G)=\{ (i_{1},j_{1}), \ldots, (i_{m},j_{m})\}$. The action of
a 1-fair edge schedule {\bf S} (specified by permutation $\pi$ of $\{1,\ldots, m\}$)
on a configuration ${\bf X}$ corresponds to multiplication of ${\bf X}$ by

\begin{equation}\label{second-eq}
\pi(S)= A_{i_{\pi[1]},j_{\pi[1]}}\cdot A_{i_{\pi[2]},j_{\pi[2]}}\cdot \ldots \cdot A_{i_{\pi[m]},j_{\pi[m]}}.
\end{equation}

An edge-schedule cannot prevent self-stabilization implies
that starting at any initial configuration, the system reaches a fixed point.
In other words, $\forall {\bf I} \in {\bf Z}_{2}^{n}$
$\exists k \in {\bf N} \ s.t. [\pi({\bf S})]^{k}\cdot {\bf I}  = {\bf 0}$.
Since the number of vectors ${\bf I}$ is finite, this is equivalent to saying that
$\exists k_1 \in {\bf N}, \forall {\bf I}, \ [\pi({\bf S})]^{k_1}\cdot {\bf X} = \textbf{0}$.
Equivalently this means that $[\pi({\bf S})]^{k}= {\bf 0}$, i.e. matrix $\pi[{\bf S}]$ is nilpotent.
Thus, what we want to show is

\begin{lemma}
For any graph $G$ there exists a schedule
{\bf S}  such that the corresponding matrix $\pi[{\bf S}]$ is not nilpotent.
\end{lemma}

Consider now an arbitrary ordering of vertices in $G$ and let $\pi$ be the
permutation corresponding to the induced lexicographic ordering of
edges of $G$ (where an edge is seen as an ordered pair, with the vertex
of lower index appearing first).

It is easy to see that

\begin{equation}\label{prod}
\Delta_{i,j}\cdot \Delta_{k,l}= \left \{\begin{array}{ll}
             \Delta_{i,l}  &\mbox{, if } j=k \\

             0  &\mbox{, otherwise. }
             \end{array}
      \right.
\end{equation}

Indeed, if $\Delta_{i,j}=(a_{m,n}^{(i,j)})_{m,n\geq 1}$, then the only wat for some element $c_{m,n}$ of
the product $\Delta_{i,j}\cdot \Delta_{k,l}$ to be nonzero is that there exists at least one term
 $d_{m,p}^{(i,j)}\cdot d_{p,n}^{(k,l)}$ that is nonzero. But this is only possible for $(i,j)=(m,p)$
and $(p,n)=(k,l)$, in other words for $m=i$, $n=l$ and $p=j=k$, which immediately yields
equation~(\ref{prod}).

Let us now consider the integer matrix $\Pi[S]$ obtained by
 interpreting equations~(\ref{first-eq}) and~(\ref{second-eq}) as
 equations {\em over integers}. Because integer addition and
 multiplication commute with taking the modulo 2 value, matrix
 $\pi[S]$ can be obtained by applying reduction modulo 2 to every
 element of $\Pi[S]$.

Let $P_{i,j}=\Delta_{i,j}+\Delta_{j,i}$.  From the definition of
matrix $\pi(S)$ in equation (\ref{second-eq}) and the definition of matrices
$A_{i,j}$ in equation (~\ref{first-eq}) we see that $\pi[S]$ is a sum
of products, each term in a product corresponding to either a
$P_{i,j}$ or to the identity matrix. Thus

\begin{equation}\label{third-eq}
\Pi[S] = I + \sum_{\emptyset \neq S\subseteq \{1,\ldots, m\}}(\prod_{k\in S} P_{i_{\pi[k]},j_{\pi[k]}}).
\end{equation}

 Consider the directed graph $\overline{G}$ obtained from $G$ by
duplicating every edge $\{i,j\}$ of $G$ into two {\em directed} edges
$(i,j)$, $(j,i)$ in $\overline{G}$. Label every edge $e\in E(G)$ by
the (unique) integer $k$ such that $e=\{i_{\pi[k]},j_{\pi[k]}\}$, and
apply the same labelling to the two oriented versions of edge $e$ in
$\overline{G}$. Then equation~(\ref{prod}) shows that nonzero products
of matrices $\Delta$ are in bijective correspondence to directed paths of
length two with increasing labels, when read from the starting to the
end node of the path.  Inductively generalizing these observations to
all sets $S$ we see that products in~(\ref{third-eq}) are nonzero
exactly when they specify a {\em directed path in $\overline{G}$} (i.e.
a path in $G$) from a vertex $k$ to a vertex $l$ with increasing
labels when read from $k$ to $l$, in which case they are equal to
$\Delta_{k,l}$.

Therefore $\Pi[S]= I+C$, where $C=(c_{i,j})$ is given by
\begin{equation}\label{fourth-eq}
c_{i,j}=  \left \{\begin{array}{ll}
             \# \mbox{ of  paths from $i$ to $j$ with increasing
labels, }
                         &\mbox{ if such paths exist} \\
              0,  &\mbox{ otherwise. }
             \end{array}
      \right.
\end{equation}

The matrix $\pi[{\bf S}]$ is, of course, obtained by reducing modulo 2
the elements of $\Pi[S]$.  A well known result in linear
algebra\footnote{justified as follows: a classical result states that
the characteristic and the minimal polynomial of a matrix have the
same roots (with different multiplicities). But it is easy to see that
the minimal polynomial of a nilpotent matrix is $x^{k}$ for some
$k\leq n$.} is that the the characteristic polynomial of a nonzero
nilpotent matrix $A$ is $x^{n}$.  Thus, one strategy to show that a
given matrix $\pi[{\bf S}]$ is not nilpotent is to make sure that for
some $p$, $0 \leq p \leq n$, the sum $s_{p}$ of its principal minors
of order $p$ is non-zero. This is equivalent to making sure that the
sum of the corresponding minors of the associated {\em integer} matrix
is odd. The proof consists of three cases:

\underline{{\bf Case (a) $G\in {\cal G}_{1}$:}} \ We will prove the following

\begin{lemma}
There exist two permutations $\sigma_{1}$ and $\sigma_{2}$ with
corresponding matrices over integers $A_{1}=\Pi[\sigma_{1}]$ and $A_{2}=\Pi[\sigma_{2}]$ such that
\[ trace(A_{2})\equiv (trace(A_{1})+1) \mbox{(mod~2)}. \]
\end{lemma}

Given Lemma 4, the proof of Lemma 3 for Case (a) follows since
matrices $\pi[\sigma_{1}]$ and $\pi[\sigma_{2}]$
cannot both be nilpotent. This is true since the $trace(A_{1})$ and $trace(A_{2})$
have different parities.
We will prove Lemma 4 using a multistep argument, combining the
conclusions of  Lemmas 3-5 below.
Consider first the following ``basic" graphs: $K_{4}$, $K_{3}\bigtriangleup K_{3}$
(the graph obtained by merging two triangles on a
common edge), and $C_{n}$, $n\geq 4$.

\begin{lemma}
The conclusion of Lemma 4 is valid for the ``basic" graphs.
\end{lemma}

\begin{proof}
By the previous result on the value of coefficients $c_{i,j}$
the value of the trace of a matrix $A$ can be easily computed from
the number of cycles with increasing labels. Also, note the following:
\begin{itemize}
\item Any cycle $C$ contributes a one to {\em at most} one
$c_{i,i}$, for some vertex $i$ appearing in $C$. This is because of the
restriction on the increasing labels, who might be verified for at most
one node of the cycle.
\item Moreover,
any triangle contributes a 1 to {\em exactly one} $c_{i,i}$.
Therefore, in considering the trace of matrix $A$ triangles add
the same quantity irrespective of permutation, and can thus be ignored.
\end{itemize}

This observation leads to a simple solution when the underlying
graph is a simple cycle $C_{n}$, $n\geq 4$, or the graph $K_{3}\bigtriangleup K_{3}$.
Note that these graphs contain a unique cycle $C$ of length atleast $4$. Consider an
ordering of the edges of this cycle, corresponding to moving around the cycle.
We will create two labellings corresponding to this ordering.
The first one assigns labels 1 to $|C|$ in this order. The
other labelling assigns labels $1,2,\ldots, |C|-2, |C|, |C|-1$ in this
order. It is easy to see that the first ordering contributes a 1 to exactly one
diagnonal element, while the second one does not contribute a 1 to any element.
Hence the traces of the corresponding matrices differ by exactly 1.

For graph $K_{4}$ we first label the
diagonal edges by 5 and 6.
There are three cycles of length $4$ in the graph $K_{4}$ --  one that uses no diagonal edges,
the other two using them both.
For the outer cycle consisting of no diagonal edges,
we consider the two orderings described in the
previous case on the outer cycle $C_{4}$. This as before shows that the traces
of the corresponding matrices differe exactly by 1.
Next, note that irrespective of the labelling
of the non-diagonal edges, the two cycles containing the diagonal edges
{\em cannot} be traversed in increasing label order, so they do not
contribute to the trace of the associated matrix.  Therefore,
the result follows for graph $K_{4}$ as well.
This completes the proof of
Lemma 3.
\end{proof}

\begin{lemma}
Let $G$ be a graph and let $G_{2}$ be a subgraph of $G$ induced by a
subset of the vertices in $G$. If the conclusion of
Lemma 3 holds for $G_{2}$, then it holds for $G$.
\end{lemma}
\begin{proof}
 Extend a permutation of the edges in $G_{2}$ to a permutation
of the edges in $G$ via a fixed labelling of the edges in
$E(G)\setminus E(G_{2})$ such that
\begin{enumerate}
\item The index of any edge with both ends in $G_{2}$ is strictly
smaller than the index of all edges not in this class.

\item The index of any edge with {\em exactly} one end in $G_{2}$
is strictly larger than the index of any edge not in this class.
\end{enumerate}

The trace of the resulting matrix is determined by the
cycles with strictly increasing labels. There are several types of
such cycles:

\begin{enumerate}
\item Cycles in $G\setminus G_{2}$. Whether such a cycle can be traversed in
increasing label order {\em does not depend on the precise labelling on
edges of $G_{2}$} as long as the conditions of the extension are
those described before.

\item Cycles containing some edges in $G_{2}$, as well as
additional edges from $G\setminus G_{2}$.
Because of the restriction we placed on the labelings, the
only such cycles that can have increasing labels
are the triangles with two vertices in $G_{2}$ and one vertex in
$G\setminus G_{2}$. Since there exist an unique way to ``read" a
triangle in the increasing order of the edge labels, their
contribution to the total trace is equal to the number of such
triangles, and {\em does not depend on the precise labeling of
edges in $G_{2}$}, as long as the restriction of the labeling is met.

\item Cycles entirely contained in $G_{2}$.
\end{enumerate}

Let now $\sigma_{1}, \sigma_{2}$ be labelings on $G_{2}$ verifying the
conclusion of Lemma 2 and $\overline{\sigma_{1}},
\overline{\sigma_{2}}$ extensions to $G$ verifying the stated
restriction. The conclusion of the previous analysis analysis is that
a difference in the parity of the traces of matrices corresponding to
the labelings $\sigma_{i}$ on $G_{2}$ directly translates into a
difference in the parity of the traces of matrices corresponding to
labelings $\overline{\sigma_{i}}$ on $G$.
\end{proof}

Finally, we reduce the case of a general graph to that of a
{\em base case} graph via the following result.

\begin{lemma}
Let $G$ be a graph that contains a cycle of length $\geq 4$ and is
minimal (any induced subgraph $H$ does {\em not} contain a cycle of
length $\geq 4$ any more).  Then $G$ is
one of the ``basic'' graphs from Lemma 3.
\end{lemma}

\begin{proof}
Let $G$ be minimal with the property that it contains a cycle of
length $\geq 4$, let $n$ be the number of nodes in $G$ and let $C$ be
a cycle of length $\geq 4$ in $G$.  Because of minimality, $C$
contains all the vertices of $G$ (otherwise $G$ would not be minimal,
since one could eliminate nodes outside $C$). Thus $C$ is a
Hamiltonian cycle. If no other edge is present then we get the cycle
$C_{n}$. Moreover, no other edge can be present unless $n=4$
(otherwise $G$ would contain a smaller cycle of length $\geq 4$ and
thus would not be minimal). In this case the two possibilities are
$K_{4}$ and $K_{3}\bigtriangleup K_{3}$.
\end{proof}

\underline{{\bf Case (b) $G\in {\cal G}_{2}$:}} \
Consider the ordering $<_{sum}$ on the edges of
$G$ so that $\{i,j\}<_{sum} \{k,l\}$ when either
\[ i+j<k+l \]
or
\[ i+j=k+l \mbox{ and } \min\{i,j\}< \min \{k,l\}. \]

In this case $c_{k,k}=0$ for every $k$, except when $k$ is the
{\em middle-index vertex} of a triangle (i.e. a triangle with vertex
 labels $i,j,k$ such that $i<k<j$).

We infer that $s_{1}=trace(A)$ is congruent (mod 2) to $n$ plus
the number of triangles in $G$, that is to $m+1$ (mod 2) (where $m$
is the number of edges of $G$).

\underline{{\bf Case (c) $G\in {\cal G}_{3}$:}}\
Consider the sum $s_{2}$ of principal minors of size
2 of $A$. In a tree there can be at most one path between two
nodes. Since we are counting paths with increasing labels, the only
way for $a_{i,j}=a_{j,i}=1$ to hold is that vertices $i$ and $j$ be
adjacent. But in this case the corresponding minor is zero. It follows
that $s_{2}$ is the number of sets of different nonadjacent vertices
in $G$, that is
\[ s_{2}= {{n}\choose {2}} - (n-1)= \frac{(n-1)(n-2)}{2} = 1 \mbox { mod 2} \]
if $n= 4k$.
\end{proof}

One might suspect that Theorem 2 (ii) extends to all graphs, thus
strenghtening the statement of Theorem 1 to 1-fair daemons. This is
not the case: Let the line graph $L_{6}$ be an instance of PDPD.  Then
{\em for all $1$-fair edge-schedulers {\bf S} and for all initial
configurations $I$ the system self-stabilizes starting at $I$.}  We
verified this statement via computer simulation, by running PDPD for
all 6! 1-fair daemons. A result that rendered this experiment
computationally feasible is the {\em state-reduction} technique
highlighted in the proof of (ii): to prove self-stabilization we only
needed to consider those initial configurations with exactly one 1. Thus
we had to run $6\times 6!$ simulations. It is an open problem to find
{\em all} graphs for which this happens. However, Theorem 2 (ii) shows that
the class of such graphs is really limited.

\section{Nonadaptive node-schedulers}

As we saw, even 1-fair edge schedulers are able to prevent self-stabilization.
What if we only allow the scheduler to choose {\em one} of the nodes ?
In this section we study the Prisoners dilemma with Pavlov dynamics when adversaries are
1-fair node-schedulers. Because only one of the nodes of the
scheduled edge is chosen by the adversary and the other one is chosen
randomly, the self-stabilization of the system is a stochastic
event.

\begin{thm}
Let $Star_{n}$ be an instance of PDPD.
Then
$S_{n}$
self-stabilizes with probability 1 against {\em any} 1-fair scheduler.
\end{thm}
\begin{proof}
One can assume, without loss of generality, that the first node to
be scheduled is the center (labeled 0) and the rest of the nodes
are scheduled in the order $1,2,\ldots, n$. Indeed, if the center
was scheduled later in the permutation of nodes, it is enough to
prove self-stabilization from the configuration that corresponds
to first running the system up to just before the center is scheduled,
and then viewing the run as initialized at the new configuration,
and with a new periodic schedule (that now starts with node 0). As
for the order in which the other nodes get scheduled, by relabelling the
nodes we may assume without loss of generality that this is $1,2,\ldots, n$.

Let $a_{0}, a_{1}, \ldots, a_{n}$ be the labels of nodes at the
beginning of the process. It is useful to first consider a deterministic
version of the dynamics in question, specified as a {\em game} between two players
\begin{itemize}
\item The first player is choosing one node to be scheduled. It is
required that the sequence of nodes chosen by this player forms a
periodic sequence $\pi$. The goal of the first player is to prevent
self-stabilization.
\item Given a node choice by the first player, the second player is
responding with a choice of the second node to be scheduled. Unlike
the first player, the sequence of nodes chosen by the second player
can potentially vary between successive repetitions of the permutation
$\pi$. The goal of the second player is to make the system converge to
state {\bf 0}.
\end{itemize}

The game above is an example of the {\em scheduler-luck games} from
the self-stabilization literature \cite{scheduler-luck}. We will
provide a strategy for the second player that (when applied) will turn any
configuration into the ``all zeros'' configuration. But a winning strategy for the second
player in the scheduler-luck game will be played with
positive probability in any round of the scheduler. Thus with probability going to one (as the
number of rounds goes to infinity) this strategy will be played at least in one round,
making the system converge to state {\bf 0}.

The crux of the strategy is to carefully use the ``partner node''
of node 0, when this is scheduled, to create a segment of nodes
$1,2, \ldots i$ (with $i$ nondecreasing, and eventually reaching
$n$) with labels zero at the beginning of a round of scheduling.

This is simple to do at the very beginning: if node 0 plays node 1
(when 0 is scheduled), then the labels of the two node will be
identical, thus when node 1 is scheduled (and plays again node 0)
the label of node 1 will be zero.

If at the beginning of a round the label of node 0 is 1, we make it
play (when scheduled at the beginning of a round) the node of smallest
positive index ($i+1$) still labelled 1. This will turn the labels of
both nodes to 0. Further scheduling of nodes $1$ to $i+1$ will not
change this, and at the end of the round, nodes $1$ to $i+1$ will still
be labelled 0.

If, on the other hand at the beginning of the round node 0 is labelled
0, we make it keep this label (and, thus, not affect the zero labels
of nodes $1$ to $1$) by making it play (when scheduled) against
another node labelled 0 (say node 1).

To complete the argument it remains to show that for any configuration
$x_{0}, \ldots x_{n}$ different from the ``all zeros'' configuration,
in a finite number of rounds we will reach a configuration where the
first case applies, and thus the length of the ``all zero'' initial
segment increases.

Indeed, assume that $x(0) = 0$ and it stays that way throughout the
process. Then, denoting by ${\bf Y_{t}} = (x(1)_{t}, \ldots x(n)_{t})^{T}$ the
labels of the nodes $1$ to $n$ at the beginning of the $t$'th round,
it is easy to see that the dynamics of the system is described by the
recurrence
\[
{\bf Y_{t+1}} = B\cdot Y_{t},
\]

with $B=(b_{i,j})$ is a matrix of order $n$ over ${\bf Z}_{2}$ specified by
\[
b_{i,j}=\left \{\begin{array}{ll}
             1  &\mbox{, if } i\geq j \\

             0  &\mbox{, otherwise. }
             \end{array}
      \right.
\]

Consider now $B$ as a matrix over {\bf Z},
rather than ${\bf Z}_{2}$. It is immediate to show by induction that
$B^{k}=(b^{(k)}_{i,j})$ given by
\[
b^{(k)}_{i,j}=\left \{\begin{array}{ll}
             {{i-j+k-1}\choose {k-1}}  &\mbox{, if } i\geq j \\

             0  &\mbox{, otherwise. }
             \end{array}
      \right.
\]

The $k$'th power over ${\bf Z}_{2}$ is obtained, of course, by
reducing these values mod 2.
In particular define $S_{t}$ to be the sum $x(1)_{t}+\ldots x(n)_{t}$. It
is easy to see that $S_{t}=x(0)_{t+1}$ thus by our hypothesis $S_{t}$ has to be zero. On
the other hand a consequence of the previous result is that
\[
S_{t}=\left[\sum_{i=1}^{n} {{n-i+t-1}\choose {t-1}}\cdot x_{i} \right ] (\mbox{  mod }2).
\]
In particular
\[
\Delta x_{t} := S_{t+1}-S_{t}= \left[\sum_{i=1}^{n-1} {{n-i+t-1}\choose
{t}}\cdot x_{i} \right ]
(\mbox{  mod }2).
\]

By induction and algebraic manipulation
we generalize this to higher order of
iterated differences
$\Delta^{k} x_{t} = \Delta(\Delta^{k-1} x_{t})$ as:
\[
\Delta^{k} x_{t} = \left [\sum_{i=1}^{n-k} {{n-i+t-1}\choose {t+k-1}}\cdot x_{i} \right ]
(\mbox{mod }2).
\]

Let $i_{0}$ be the smallest index such that $x(i_{0}) = 1$. Then, by
the previous relation
\[
\Delta^{n-i_{0}} x_{t}= \left [\sum_{i=1}^{i_{0}} {{n-i+t-1}\choose
{t+k-1}}x_{i}\right ] = {{n-i_{0}+t-1}\choose {n-i_{0}+t-1}}\cdot
x_{i_{0}} = x_{i_{0}} = 1 \mbox{ (mod 2).}
\]

But this contradicts the fact that $S_{t}=0$ for every value of $t$
and completes the proof.
\end{proof}

A 1-adaptive scheduler keeps repeating the nodes to
choose according to a fixed permutation. Thus, for a fixed scheduler
and interaction graph we can talk about the probability of
stabilization in the limit. Also, for a given fixed scheduler, the
event that this limit is one is a deterministic statement.
Consequently we can talk of the probability that this event
happens when the interaction graph is sampled from a class of random
graphs. As noted, for a random scheduler the condition that $G$
has no isolated vertices is necessary and sufficient to guarantee
self-stabilization with probability 1. This is also true for the
adversarial model in the case of non-adaptive (1-fair) daemons. This is in
case with the case of an edge
daemon, when even non-adaptive daemons could preclude
stabilization.

\begin{thm}
Let $G$ be an instance of PDPD such that $G$
has no isolated vertices. Then
for {\em any} 1-fair node-scheduler and any initial configuration the system $G$
reaches state {\bf 0} with probability 1.
\end{thm}

The results of Theorem 5 should be contrasted with
the corresponding result for edge schedulers, for which, as we showed,
even non-adaptive daemons could preclude  stabilization.

The proof consists of the following three components:
\begin{enumerate}
\item our earlier result that guarantees a
winning strategy for scheduler-luck game associated to the dynamics when
the underlying graph is $S_n$,
\item the partition of a spanning forest of $G$ into node disjoint
stars and
\item  the fact that the existence of such a winning strategy is a
{\em monotone graph property} w.r.t to edge insertions.
This is formally stated in the following
\end{enumerate}

\begin{lemma}
Suppose $H$ is a graph such that a winning strategy $W$ exists for
the scheduler-luck game on a graph $H$. Let $e\not \in E(H)$, and let
$L=H\cup\{e\}$. Then $W$ is also a winning strategy for any graph $L$.
\end{lemma}

\begin{proof}
Given any node choice by the first player, the second player can
choose the corresponding node according to strategy $W$ (thus never
scheduling the additional edge $e$). The outcome of the game is,
therefore, identical on $H$ and $L$.
\end{proof}

\noindent
{\bf Proof of Theorem 5.}
By Lemma 6 it is enough to show the existence
of a winning strategy for the second player in the scheduler-luck game on a graph
$G$, when $G$ is a tree.
We decompose tree $G$ into a set $\{S_{1}, \ldots, S_{p}\}$ of
node-disjoint stars as follows.
\begin{itemize}
\item
Root $G$ at an arbitrary node $r$.
\item
Consider the star formed by the root and its children.
Call it $S_1$.
\item
Remove the nodes in $S_1$ and all edges with one end point
incident on nodes in $S_1$
\item
Recursively apply the procedure on each forest created by the above
operation.
\end{itemize}

Now consider a 1-fair schedule $\pi$ on graph $G$, corresponding to a
strategy of the first player in the schedule-luck game on $G$. For
every star $S_{i}$, the projection $\pi_{i}$ of the schedule on the
nodes of $S_{i}$ (that amounts to only considering scheduled nodes
that belong to $S_{i}$) specifies a 1-fair schedule on
$S_{i}$. According to Theorem 5, the second player has a
winning strategy $W_{i}$ for the scheduler-luck game on $S_{i}$ when
the first player acts according to the schedule $\pi_{i}$.

Next, we devise a strategy $W$ for the scheduler-luck game on graph $G$, by
by ``composing'' the winning strategies $W_{i}$. Specifically if the
node chosen by the first player belongs to star $S$, strategy $W$
will employ $W_{i}$ to choose the corresponding second node. Since on
each star $S_{i}$ the labels of the node will eventually be {\bf 0},
$W$ is a winning strategy for the second player in the scheduler-luck
game on $G$. \qedbox

\section{Adaptive node schedulers}

Nonadaptive schedulers could not preclude self-stabilization.  In
contrast, as the following theorem shows, $3$-fair nonadaptive
node-schedulers are still powerful enough to preclude
self-stabilization {\em with complete certainty}, and so are {\em $2$-fair
 adaptive}\footnote{obviously, there are no 1-fair adaptive schedulers.} schedulers.

\begin{thm}
The following are true:
\begin{enumerate}
\item Let the star graph $S_n$ ($K_{1,n}$) be an instance of PDPD.  Then
there exists an initial configuration $I$ and a $3$-fair nonadaptive
scheduler that precludes self-stabilization on $Star_n$ starting in
$I$.
\item Let the triangle $K_{3}$ be an instance of PDPD. Then there exists
an initial configuration $I$ and a $2$-fair {\em adaptive} scheduler that
precludes self-stabilization on $K_{3}$ starting in
$I$.
\end{enumerate}
\end{thm}

\begin{proof}
\begin{enumerate}
\item Consider the star graph $S_n$ ($K_{1,n}$), with
 the center labeled 0 and the rest of the nodes labeled $1,2,\ldots, n$.
We have to provide an example of a $3$-fair scheduler
that precludes self-stabilization on some initial configuration. This
initial configuration has two 1's, at nodes 1 and 2. The scheduler
repeats the schedule $[0, 1, 1, 3, 4, \ldots n-2, 2, 1, n-1,
n-1]$. After the scheduling of 0 1 1 the effect is that both nodes
have label 0. Thus the scheduling of nodes $3,4, \ldots, n-2$ does not
change any label. With node 2 the label of node 0 will change to $1$,
thus changing in the next step the label of node 1 back to
$1$. Finally scheduling the node $n-1$ twice turns back the label of
node 0 to 0, thus yielding the initial configuration.
It is easy to see that the scheduler is
$3$-fair.
\item Start with configuration $I$ consisting of all ones. The scheduler will adaptively
schedule the nodes, in sequences of three, so that at the end of such
a {\em 3-block} the system is guaranteed to be in configuration $I$
again. Figure~\ref{as} describes the strategy of the scheduler, assuming
that node 1 is scheduled first.

\begin{figure}
\begin{center}
\includegraphics*[width=8cm]{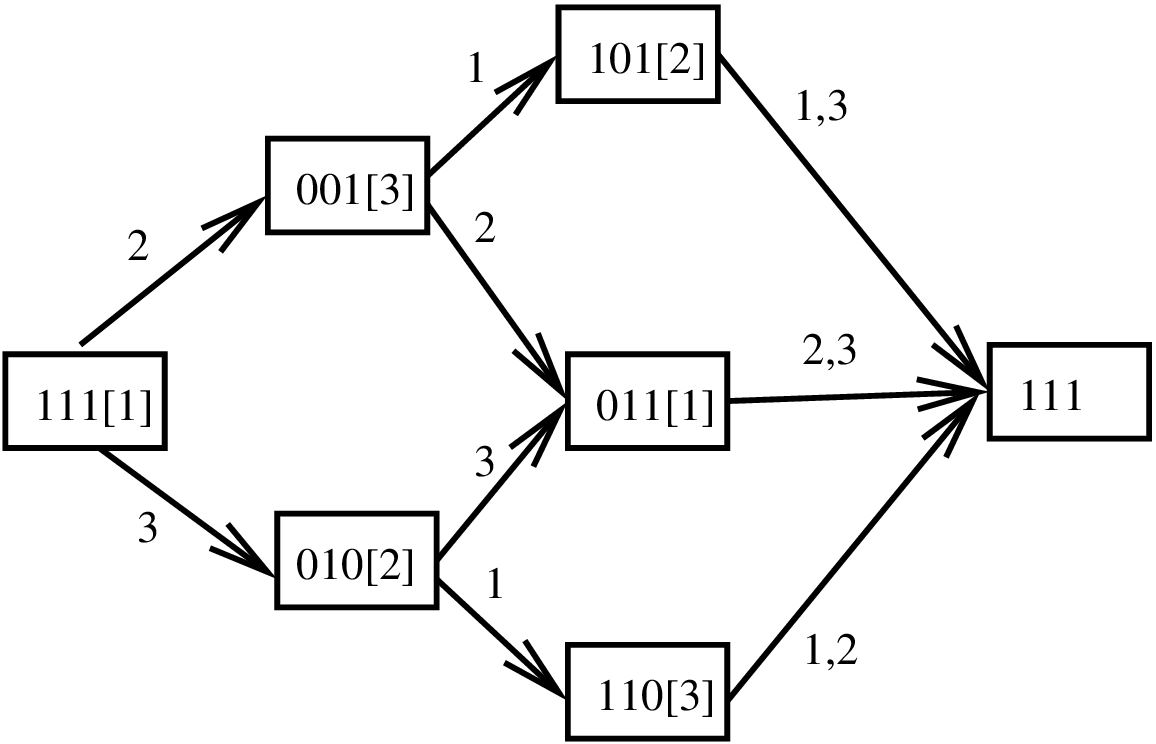}
\caption{A round of the 2-fair adaptive scheduler}
\label{as}
\end{center}
\end{figure}

Elements in the rectangle represent the
state of the system, followed by the scheduled node (in square brackets).
choices. Labels on the edges represent the possible probabilistic choices
of the partner node, with multiple (inconsequential) choices separated by a comma.
Note that the scheduler has similar
strategies if one of the nodes 2,3 is scheduled first. Also, note that a
3-block consists of either a permutation of a node, or two nodes, with
the initial and the final node in the block being identical. The
scheduler proceeds now to create an infinite schedule consisting of 3-blocks
according to the following rule:
\begin{enumerate}
\item If a given block $B$ is a permutation then start the next block with the same starting
element as $B$.
\item Otherwise if the given block $B$ is missing node $z$, start the next block by first
scheduling $z$.
\end{enumerate}

It is easy to see that the scheduler we constructed is 2-adaptive and precludes self-stabilization.
\end{enumerate}
\end{proof}

Although formal definition of the probability of self-stabilization is
more complicated in this case, we can talk of the probability of
self-stabilization for adaptive daemons as well. However, as we have
seen, the result of Theorem 5 is {\em no longer true}: on stars,
1-fairness is stronger than $2$-fair adaptive scheduling.  It would
seem that this result shows that nonadaptiveness is important for
self-stabilization. However, we will see that the class of network
topologies where this happens is reasonably limited. Indeed, we next
study self-stabilization on Erd\H{o}s-Renyi random graphs $G(n,p)$.
We will choose $p$ in such a way that with high probability a random
sample from $G(n,p)$ has no isolated vertices\footnote{A random sample
from $G(n,p)$ has no isolated vertices with probability $1-o(1)$ when
\cite{janson-random-graphs} $np-\log{n} \rightarrow \infty$.}. In
other words, we require that necessary condition on the topology of
$G$ holds with probability $1-o(1)$. Call a graph $G$ to be {\em good}
if for {\em any scheduler of bounded fairness} and any starting
configuration $I$, $G$ starting at $I$ converges to {\bf 0} with
probability $1-o(1)$, as the number of steps goes to infinity.

\begin{thm}
Let $p$ be s.t. $np-\log{n} \rightarrow \infty$.
Then with probability $1-o(1)$ a random graph $G\in G(n,p)$ is good.
\end{thm}

Of course, a natural question is whether such a weakening of the
original result, from any graph topology satisfying a given condition
to a generic random graph satisfying the same condition, is
reasonable. We are, however, not the first ones to propose such an
approach. Indeed, except for a handful of cases the network that a
given social dynamics takes place on is not known in its
entirety. Instead, a lot of recent work (see
e.g. \cite{vespignani-internet,structure-dynamics-networks,durrett-graphs}
for presentations) has resorted to the study of  {\em
generic} properties of random network models that share some of the observable
properties of a {\em fixed} network (such as the Internet or the World
Wide Web.

\begin{proof}
The plan of the proof is similar to that for 1-fair schedulers.
Define a {\em round} of a $b$-fair scheduler to consist of a
consecutive sequence of \mbox{$b(n-1)+1$} steps.

\begin{enumerate}
\item We prove that for graphs from a class ${\bf B}$ of ``base
case'' graphs  (Lemma 7 below)
the second player has a winning strategy in the
scheduler-luck game associated to any scheduler of bounded fairness,
where the games corresponds to a finite number of rounds of the
scheduler.
\item We use the monotonicity of the existence of a strategy.
\item We show that with probability $1-o(1)$ the vertices of a
random sample graph $G$ from the graph process can be partitioned such
that all the induced subgraphs are isomorphic with one graph in $G$.
\end{enumerate}


\begin{lemma}
The following are true:
\begin{enumerate}
\item Let $G$ be a graph with a perfect matching. Then for any
scheduler $S$ of bounded fairness, and any initial configuration on
$G$, the second player has a winning strategy for the scheduler-luck
game corresponding to one round of the scheduler.
\item Let the line $L_{n}$, $n \geq 6$ be an instance of PDPD.
Then for any node-scheduler {\bf S} of bounded fairness and every
initial configuration $I$, the second player has a winning strategy in the
scheduler-luck game associated with two consecutive rounds of {\bf S}.
\end{enumerate}
\end{lemma}

\begin{proof}
\begin{enumerate}
\item A perfect matching $M$ of $G$ specifies a
winning strategy in a scheduler-luck game: each node plays (when
scheduled) against its partner in $M$. Since every node is scheduled at least once
in a round of scheduling, every edge of $M$ is played at least twice. Therefore,
irrespective of the initial configuration, the final configuration is {\bf 0}.

\item The lemma only needs to be proved in the case when $n$ is odd (in the
other case the strategy based on perfect matchings applies). In this
case the winning strategy is specified as follows:
\begin{enumerate}
\item In the first round  turn the leftmost $L_{4}$ portion of $L_{n}$ into the all zero
state by playing the matching based winning strategy.
\item In the second round nodes in the leftmost $L_{3}$ will only choose to play against each
other when scheduled, thus remaining at 0. The remaining nodes form a graph isomorphic to
$L_{n-3}$, and in this round we use the perfect matching based strategy for this graph.
\end{enumerate}
\end{enumerate}
\end{proof}

We now note now that the statement of Lemma 6 extends to
scheduler-luck games associated to node daemons of {\em bounded}
fairness. The proof is similar (a strategy for the game on $G$ is also
a strategy for the game on a graph with a larger set of edges).
Theorem 7 immediately follows if $n$ is even:
a classical result in random graph theory (see
e.g. \cite{janson-random-graphs} pp. 82-85) asserts that with
probability $1-o(1)$ $G$ will have a perfect matching.

To complete the proof of Theorem 7 we only need to deal with
the case when $n$ is odd. For a graph $G$ and a set of vertices $V$
denote by $G|_{V}$ the subgraph induced by vertex set $V$.

\begin{lemma}
With probability $1-o(1)$ $G$ can be partitioned into
$V=V_{1}\cup V_{2}$ such that:
\begin{enumerate}
\item $G|_{V_{1}}$ contains $L_{7}$ as an edge-induced subgraph.
\item $G|_{V_{2}}$ is a graph with a perfect matching.
\end{enumerate}
\end{lemma}

Theorem 7 follows from Lemma 8,
since for both $G|_{V_{1}}$ and $G|_{V_{2}}$ the second player has a
winning strategy in the scheduler-luck game. A winning strategy for
the corresponding game on $G$ proceeds by using the winning strategy
for $G|_{V_{1}}$, when the scheduled node is in $V_{1}$ and the
winning strategy for $G|_{V_{2}}$ when the scheduled node is in
$V_{2}$.

 The proof of Lemma 8 goes along lines similar to that of
the proof of the existence of a perfect matching in a random grapg(see
\cite{janson-random-graphs} pp. 82-85). The first step is to show that
w.h.p. $G$ does not contain a set of distinct vertices $x_{0}, x_{1},
x_{2}, x_{3}, x_{4}$ such that:
\begin{itemize}
\item $deg_{G}(x_{0})= deg_{G}(x_{4})=1$.
\item For every $i=\overline{0,3}$, $x_{i}$ and $x_{i+1}$ are adjacent.
\end{itemize}

This is easy to see, since the expected number of such structures is
$O(n^{5}p^{4}(1-p)^{2(n-1)}) = O(n^{5}p^{4}e^{-2np})=o(1)$,  since
$p=\Theta(\frac{\log (n)}{n})$.

Consider now a random graph $G$, conditioned on not containing such a
structure, and a vertex $x_{0}$ in $G$ of degree one. Since this
information only exposes information on the edges with one endpoint at
$x_{0}$, with probability $1-o(1)$ graph $H=G\setminus \{x_{0}\}$ has
a perfect matching. Let $x_{1}$ be the node in $G\setminus \{x_{0}\}$
adjacent to $x_{0}$, and let $x_{2}$ be the node matched to $x_{1}$ in
$H$. With probability $1-o(1)$ $x_{2}$ has another neighbor $x_{3}$ in
$H$, (otherwise $G$ would contain a {\em cherry}, i.e. two vertices of
degree one at distance exactly 2 in $G$ (see Figure~\ref{cherry}). But
(see \cite{janson-random-graphs} pp. 86) a random sample from $G(n,p)$
only contains a cherry with probability $o(1)$.  Let $x_{4}$ be the
node $x_{3}$ is matched to in $H$. Again, with probability $1-o(1)$
$x_{4}$ has a neighbor $x_{5}$ in $H$ different from $x_{3}$
(otherwise the five vertices $x_{0}$ to $x_{4}$ would form a structure
we have conditioned on not occurring in $G$). Finally let $x_{6}$ be
the node matched to $x_{5}$ in $H$. Then the restriction of $G$ to the
set $V_{1}=\{x_{0}, \ldots, x_{6}\}$ contains a copy of $L_{7}$, and
$G$ restricted to $V_{2}=V\setminus V_{1}$ contains a perfect matching
(induced by the perfect matching on $H$).

\begin{figure}
\begin{center}
\includegraphics*[width=4cm]{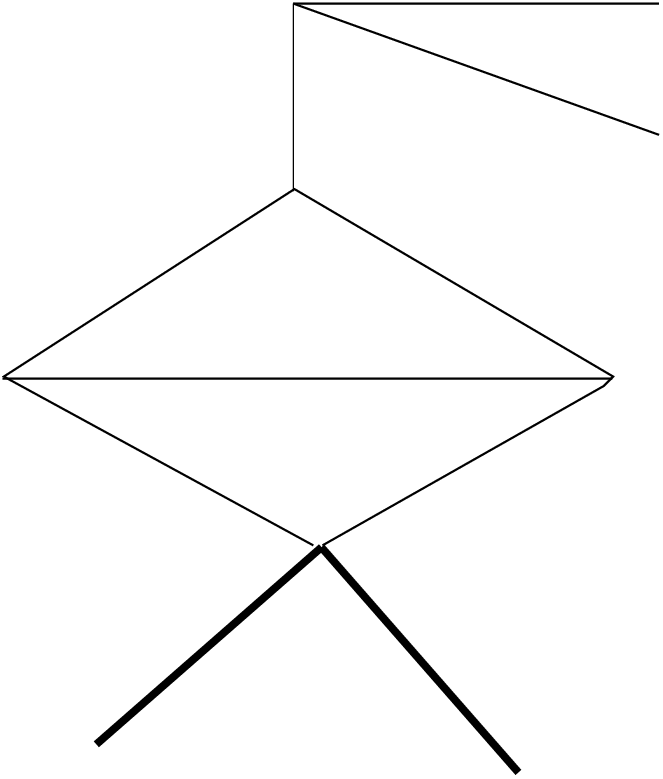}
\caption{A {\em cherry} in a graph (with bold lines)}
\label{cherry}
\end{center}
\end{figure}

\end{proof}

\section{\label{times-sim}Speed of convergence}

The previous theorems have shown that results concerning convergence
to a fixed point can be studied in (and extend to) an adversarial
framework. Perhaps what is not preserved as well in the adversarial
framework is results on the {\em computational efficiency} of
convergence to equilibrium. Such results include, for instance, the
above mentioned $O(n \log n)$ bound of \cite{ipd:colearning}. The
proof of this theorem displays an interesting variation on the idea of
a potential function. It uses such a function, but in this case the
value of the function only diminishes ``on the average'', rather than
for {\em every} possible move. Therefore bounding the convergence time
seems to critically use the ``global'' randomness introduced in the
dynamics by random matching, and does {\em not} trivially extend to
adversarial versions. On the other hand, the proof of Theorem 4 only
guarantees an exponential upper bound on expected convergence time.

We have investigated experimentally the convergence time on $C_{n}$
for some classes of 1-fair schedulers (permutations). Some of our
results are presented in Figure~\ref{time}, where we present the
average number of {\em rounds}, rather than steps, over 1000 samples
at each point.  The symbol {\bf id} denotes the identity permutation
$(1 2 \ldots n)$, {\bf p3} is the permutation $\sigma[i]=3i\mbox{(mod
$n$)}$, {\bf (13)} refers to permutations with pattern $(13245768
\ldots)$, and {\bf rd} refers to the maximum average number of rounds,
taken over 10 {\em random} permutations. In all cases the convergence
time is consistent with the above-mentioned $O(n \log n)$ result.

\begin{figure}
\begin{center}
{\small
\begin{tabular}{|c|c|c|c|c|c|c|}\hline \hline
$\pi|n$ & 4 & 8 & 16 & 32 & 64 & 128 \\
\hline \hline
{\bf id} & 2.486 & 4.225 & 6.401 & 8.33 & 10.498 & 13.135 \\
\hline
{\bf p3} & 2.469 & 4.039 & 5.807 & 7.662 & 9.639 & 11.718 \\
\hline
{\bf rd} & 2.289 & 4.499 & 6.527 & 8.781 & 11.161 & 14.151 \\
\hline
{\bf (13)} & 2.168 & 4.656 & 7.069 & 9.837 & 12.653 & 14.859 \\
\hline\hline
$\pi|n$ & 256 & 512 & 1024 & & & \\
\hline\hline
{\bf id} & 16.091 & 17.954 & 20.331 & & & \\
\hline
{\bf p3} & 14.323 & 16.054 & 19.826 & & &  \\
\hline
{\bf rd} & 17.342 & 20.518 & 22.336 & & & \\
\hline
{\bf (13)} & 18.504 & 20.346 & 20.392& & &  \\
\hline
\end{tabular}
}
\caption{Rounds on $C_{n}$ under 1-fair scheduling.}

\label{time}
\end{center}
\end{figure}

We are unable to obtain such a result (and leave it as an interesting
open problem)\footnote{A promising approach is outlined in
\cite{messika-podc05}}. It is even more interesting to study the dependency
of the mixing time of the dynamics \cite{ipd:colearning} on the
underlying network topology.  While there are superficial reasons for
optimism (for some models in evolutionary game theory,
e.g. \cite{morris-contagion}, the impact of network topology on the
convergence speed of a given dynamics is reasonably well understood),
the reader is directed to \cite{mossel-ipd} (especially the concluding remarks) for a discussion
on the difficulties of connecting network topology and convergence speed
for the specific dynamics we study.

\section{Discussion of Results and Conclusions}

We have advocated the study of evolutionary game-theoretic models under adversarial
scheduling, similar to the ones in the theory of self-stabilization.
As an illustration we studied the Iterated Prisoners'Dilemma with
the win-stay lose-shift strategy.

Our results are an illustration of the adversarial approach as follows:

\begin{enumerate}
\item {\bf Start with some result $P$, valid under random scheduling.} The original statement is presented in Subsection 2.2.
\item {\bf Identify several structural properties of a random scheduler that impact the validity of $P$.}
The random scheduler is
\begin{itemize}
\item {\em fair} more precisely $O(n\log n)$ fair w.h.p. by the Coupon Collector Lemma.
\item {\em endogeneous}, since the next edge to be scheduled is not fixed in advance.
\item {\em nonadaptive}, since the next edge to be scheduled does not depend on the configuration of
the system.
\end{itemize}
\item {\bf Identify those properties (or combinations of properties) that are
 {\em necessary/sufficient} for the validity of $P$.}

Theorem 1 shows that fairness is a {\em necessary condition}  for the
extension of the original result to adversarial settings. Next, the definition
of node and edge schedulers illustrates another important property of
random schedulers: {\em endogeneity of agent interactions}: an edge
scheduler completely specifies the dynamics of interaction. In
contrast, node schedulers provide perhaps the weakest possible form of
endogeneity: the underlying social network is still fixed, but the
agents can choose a neigbor among his neighbors to interact with (or
simply play a random one).

Theorems 4 and 5 show that, in contrast with the case of edge
schedulers, even this limited amount of endogeneity is sufficient to
recover the original result for random scheduler.  Moreover, the
proofs illuminate the role of endogeneity, that was somehat obscured
in the (trivial) original proof that the Pavlov dynamics (under random
matching) converges with high probability to the ``all zeros"
fixed-point. This proof implicitly relies on the fact that from every
state there exists a sequence of ``right" moves, that ``funnels" the
system towards the fixed point.  For {\em node} schedulers the
existence of a such a set of right moves is proved by explicit
construction and is more difficult in the adversarial setting. The
existence of such a set of moves is precisely what exogeneous choice
of agents is able to preclude.

\item

{\bf Correspondingly, identify those properties that are inessential
to the validity of $P$. In the process one can reformulate (if needed)
the original statement in a way that makes it more robust.}

Theorem 6 shows that, if we allow schedulers to be adaptive, then {\em
network topology} becomes important, and can invalidate the original
result in an adversarial setting. However adaptiveness (or,
equivalently, the amount of fairness) is {\em inesential} if we
require the convergence result to only hold {\em generically} with
respect to the class of network topologies described by
Erd\H{o}s-Renyi random graphs.
\end{enumerate}

The results we proved also highlight a number of techniques from the
theory of self-stabilization that might be useful in developing a
general theory:

\begin{itemize}
\item the concept of scheduler-luck game.
\item composition of strategies by partitioning  the
interaction topology.
\item monotonicity and ``generic preservation'' via threshold properties.
\end{itemize}

Obviously, a reconsideration of more central game-theoretic models
under adversarial scheduling is required (and would be quite interesting).

\section*{Acknowledgments.}
 This work has
been supported by the Romanian CNCSIS through a PN II/Parteneriate Grant, by the U.S.\ Department of Energy under contract
W-705-ENG-36 and  Los Alamos National Laboratory through the LANL
LDRD program, and by by NSF
Grant CCR-97-34936.

\bibliographystyle{alpha}
\bibliography{/home/gistrate/bib/bibtheory}

\end{document}